\documentclass[12pt,preprint]{aastex}

\shorttitle{The plasma structure of the Cygnus Loop}
\shortauthors{Uchida et al.}

\begin{document}


  \title{The Plasma Structure of the Southwestern Region \\ 
  of the Cygnus  Loop with the XMM-Newton Observatory}


\author{Hiroyuki Uchida\altaffilmark{1}, Hiroshi Tsunemi\altaffilmark{1}, Satoru Katsuda\altaffilmark{1} \altaffilmark{2} and Masashi Kimura\altaffilmark{1}}
%\affil{Department of Earth and Space Science, Graduate School of
%  Science, Osaka University, Toyonaka, Osaka 560-0043, Japan}
\email{uchida@ess.sci.osaka-u.ac.jp, tsunemi@ess.sci.osaka-u.ac.jp, katsuda@ess.sci.osaka-u.ac.jp, mkimura@ess.sci.osaka-u.ac.jp}


\altaffiltext{1}{Department of Earth and Space Science, Graduate School of Science, Osaka University, Toyonaka, Osaka 560-0043, Japan}
\altaffiltext{2}{NASA Goddard Space Flight Center, Code 662, Greenbelt, MD 20771}


\begin{abstract}
We observed the southwestern region of the Cygnus Loop in two pointings with \textit{XMM-Newton}. The region observed is called the ``blow-out'' region that is extended further in the south. The origin of the ``blow-out'' is not well understood while it is suggested that there is another supernova remnant here in radio observation. To investigate the detail structure of this region in X-ray, we divided our fields of view into 33 box regions. The spectra are well fitted by a two-component nonequilibrium ionization model. The emission measure distributions of heavy elements decrease from the inner region to the outer region of the Loop. Then, we also divided our fields of view into 26 annular sectors to examine the radial plasma structure. Judging from metal abundances obtained, it is consistent with that the X-ray emission is the Cygnus Loop origin and we concluded that high-$kT_{e}$ component ($\sim$0.4\,keV) originates from the ejecta while low-$kT_{e}$ component ($\sim$0.2\,keV) is derived from the swept-up interstellar medium. The flux of low-$kT_{e}$ component is much less than that of high-$kT_{e}$ component, suggesting the ISM component is very thin. Also, the relative abundances in the ejecta component shows similar values to those obtained  from previous observations of the Cygnus Loop. We find no evidence in X-ray that the nature of the ``blow-out'' region originated from the extra supernova remnant. From the ejecta component, we calculated the masses for various metals and estimated the origin of the Cygnus Loop as the core-collapse explosion rather than the Type Ia supernova.
\end{abstract}

\keywords{ISM: abundances --- ISM: individual (Cygnus Loop) ---
  supernova remnants --- X-rays: ISM}

\section{Introduction}\label{sec:intro}
The Cygnus Loop is one of the brightest Supernova Remnant (SNR) in the
X-ray sky. Its age is estimated to be $\sim$\,10,000 yrs \cite{Blair05}. Since the distance is comparatively
close to us (540\,pc; Blair et al. 2005), the apparent size is quite large
($2.5^\circ\times3.5^\circ$; Levenson et al. 1997), which enables us to study the plasma structure of the Loop in detail.

Although the \object{Cygnus Loop} is an evolved SNR, a hot plasma is still confined inside the Loop \cite{Hatsukade90}.
Miyata et al. (1998) observed the Loop with the \textit{Advanced Satellite for Cosmology and Astrophysics} (ASCA), and detected the strong  highly-ionized Si-K, S-K line and Fe-L line near the center of the Cygnus Loop. They concluded that the hot plasma, a ``fossil'' of the supernova explosion, left in the core of the Loop. 
Tsunemi et al. (2007) (hereafter TKNM07) observed the Cygnus Loop along the diameter from the northeast (NE) to the southwest (SW) with \textit{XMM-Newton} and studied the radial plasma structure. 
From the spectral analysis, they showed that the Cygnus Loop consists of two component plasma. 
They concluded that the low-$kT_e$ component originating from the interstellar medium (ISM) surrounds the high-$kT_e$ component originating from the ejecta. 
In addition, they measured the metal abundances of the high-$kT_e$ component and showed the metal distribution of the ejecta. 
The results indicate that the abundances are relatively high ($\sim$5 times solar) and each element is nonuniformly distributed: Si, S and Fe are concentrated in the inner region while the other elements such as O, Ne and Mg are abundant in the outer region. 
They also estimated the progenitor star's mass to be 15 M$_\odot$.

The Cygnus Loop is a typical shell-like SNR; this structure is thought to be generated by the cavity explosion \cite{Levenson97}.
The Cygnus Loop is almost circular in shape, however, we can see some breakout in the SW. 
It is called the ``blow-out'' region \cite{Aschenbach99}.
The origin of the ``blow-out'' is not well understood. 
Aschenbach \& Leahy (1999) have explained this extended structure as a breakout into a lower density ISM. 
On the other hand, Uyaniker et al. (2002) suggested the existence of a secondary SNR (named G72.9-9.0) in the south from a radio observation and some other radio observations support this conclusion (Uyaniker et al. 2004; Sun et al. 2006).

Our observations were performed in a direction from the Cygnus Loop center toward the south ``blow-out'' region. 
In this paper, we report the result of the spectral analysis and discuss about the plasma structure of this region.

\section{Observations}
\begin{figure}
  \begin{center}
    \includegraphics[width=80mm]{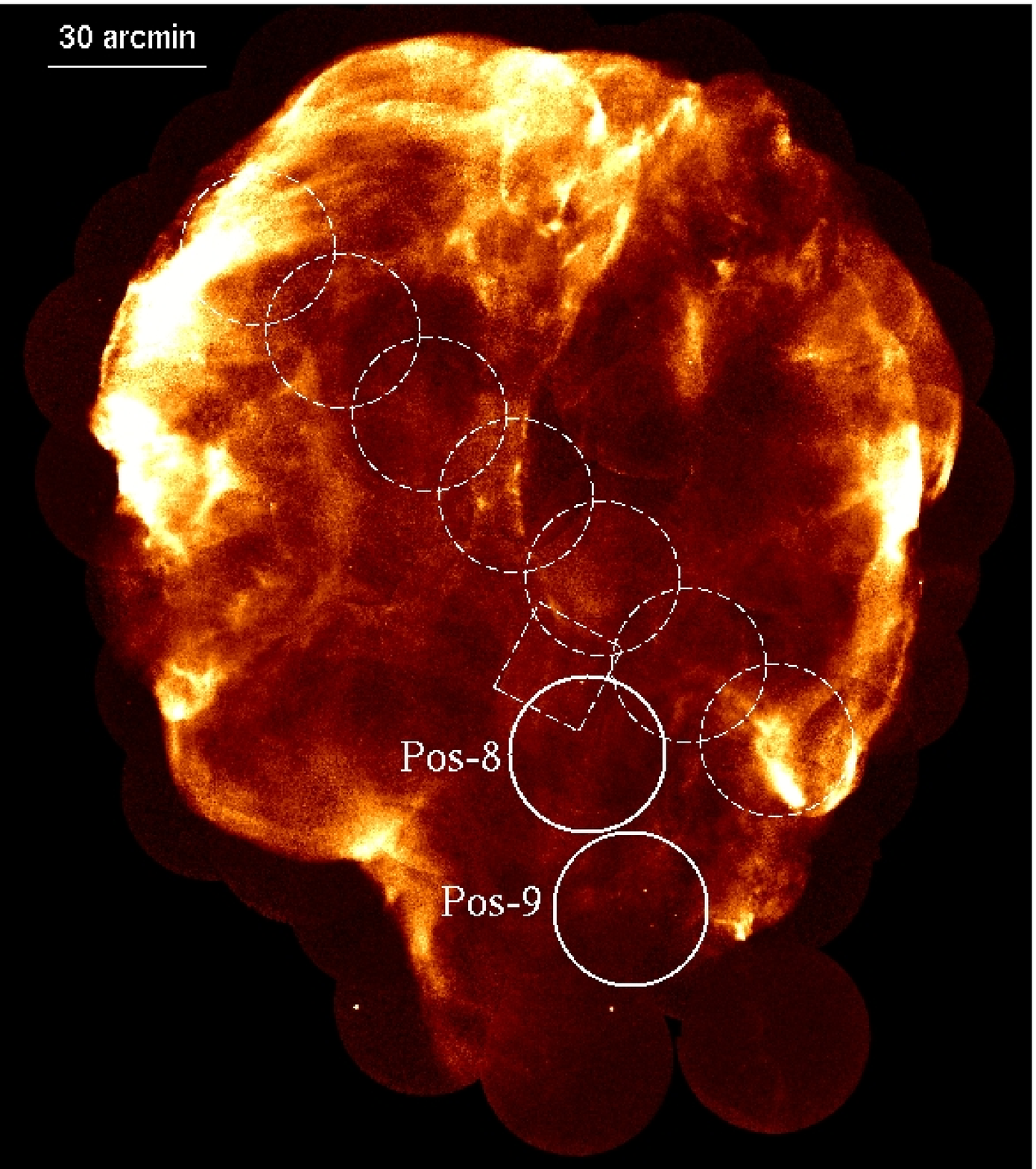}
    \includegraphics[width=65mm]{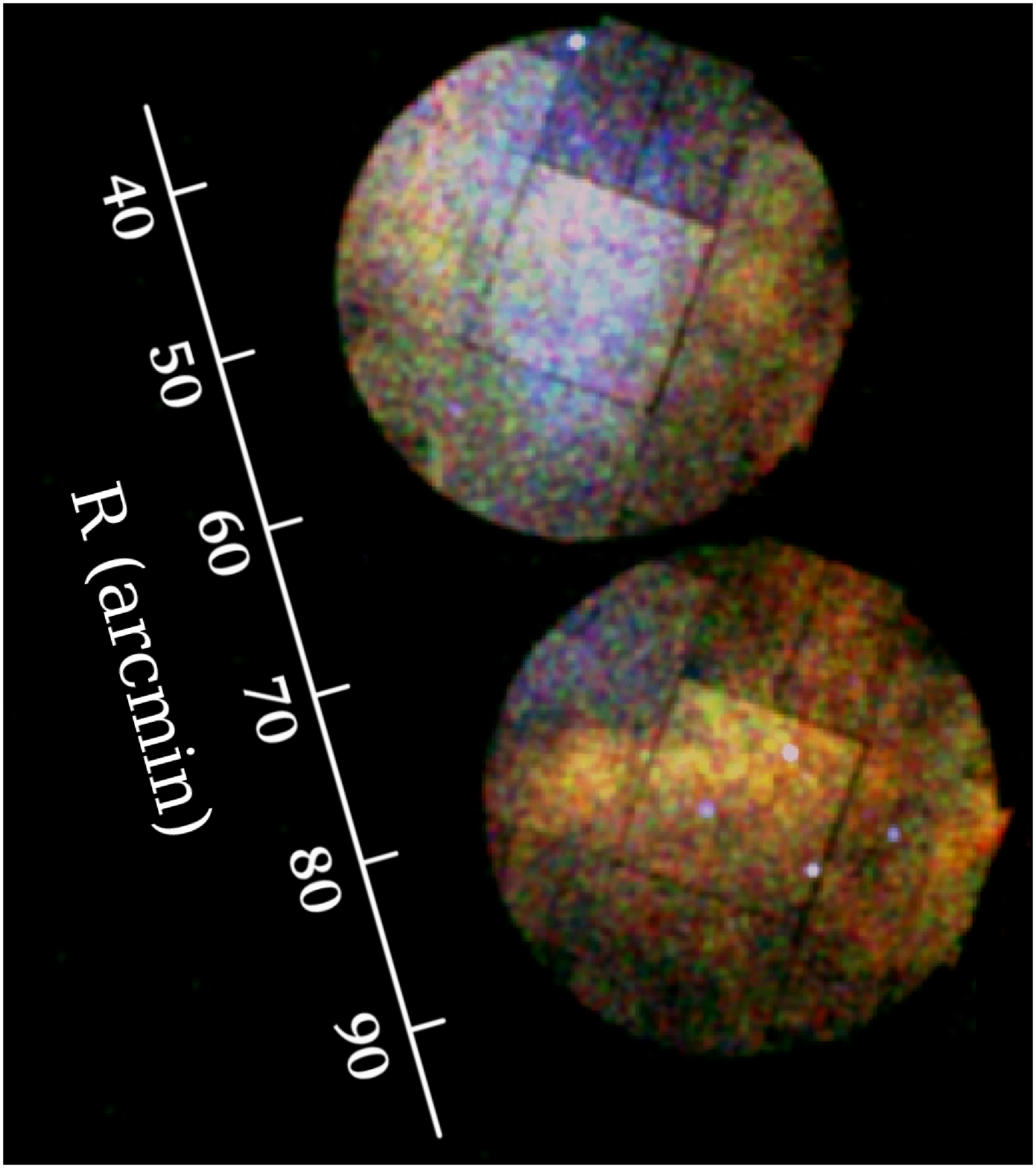}  
  \end{center}
  \caption{\textit{Left}: X-ray surface brightness map of the Cygnus Loop obtained with the \textit{ROSAT} High Resolution Imager (HRI). The dotted circles and solid circles represent the FOV of TKNM07 and our observations (Position-8 and 9), respectively. The dotted square represents the FOV of \textit{Suzaku} observation named P16 \cite{Katsuda08-2}. \textit{Right}: Three-color X-ray image for Pos-8 and 9 using EPIC MOS 1 and 2 data.}\label{fig:HRI}
\end{figure}

We performed two pointing observations of the SW region of the Cygnus Loop with the \textit{XMM-Newton} observatory. TKNM07 have observed from the NE to the SW along the diameter. 
Then, we intended to expand our observation southward from the center. 
Figure \ref{fig:HRI} left panel shows the X-ray surface brightness map of the Cygnus Loop obtained with \textit{ROSAT} High Resolution Imager (HRI) on which we showed our Field of View (FOV) of the EPIC MOS by solid white circles. 
We call the north observation for Position-8 (Pos-8) and the south observation for Position-9 (Pos-9). 
If there exists a secondary SNR in SW as Uyaniker et al. mentioned, our whole FOV overlaps the SNR whose center is roughly located at the south in Pos-9. 
Figure \ref{fig:HRI} right shows a three-color X-ray image of our FOV using \textit{XMM-Newton} EPIC MOS 1 and 2 data after correcting for exposure and vignetting effects. 
Red, green and blue correspond to the energy ranges of 0.3-0.5\,keV, 0.5-0.7\,keV and 0.7-3.0\,keV, respectively. 
Figure \ref{fig:MOS} shows the MOS broad-band image for the 0.3-3\,keV range. 
The white X shows the center of the G72.9-9.0 estimated by Uyaniker et al. (2002).

Both observations were performed on May 13 2006, during the \textit{XMM-Newton} AO-5 observing cycle. 
The total exposure time was both $\sim$10\,ks.
In order to exclude the background flare events, we considered the time intervals when the count rates were high as flare events, and eliminated them from our analysis.
The pn data were almost unusable due to the flare. Therefore we only used the data obtained with the EPIC MOS for our analysis.
The effective exposure times for each observation were 6.5\,ks (Pos-8) and 3.6\,ks (Pos-9), respectively.
Both data were taken using the medium filters and the prime full-window mode.
All the data were processed with version 7.1.0 of the \textit{XMM} Science Analysis System (SAS). 
For the background subtraction, we employed a blank-sky observations prepared by Read \& Ponman (2003).
The point-like sources were excluded using the SAS task edetect chain for the spectral analysis. As a result, one and four point-like sources are detected in Pos-8 and Pos-9, respectively. Two of the point sources in Pos-9 were observed by Miyata et al. (2001) and named AX J2049.6+2939 and AX J2050.0+2914, respectively.

\section{Spectral Analysis}\label{sec:specana}
\subsection{Two-component VNEI Model}\label{sec:vnei}
\begin{figure}
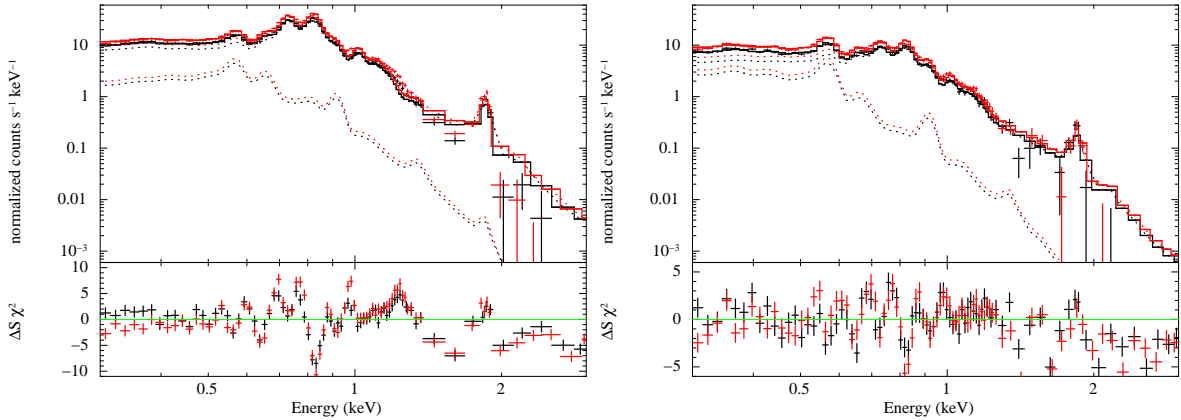

  \begin{center}
    \includegraphics[width=55mm,angle=-90]{f3a.eps}
    \includegraphics[width=55mm,angle=-90]{f3b.eps}
 \end{center}
  \caption{\textit{Left}, MOS 1 (black) and MOS 2 (red) spectrum for Pos-8 which are summed over the entire FOV. The best-fit curves are shown as solid lines. The dotted lines show individual component of the model. The lower panel shows the residual. \textit{Right}, same as the left panel, but for Pos-9. }\label{fig:spec}
\end{figure}

Figure \ref{fig:spec} shows the spectra for Pos-8 and 9 summed over the entire FOV. We can see some emission lines such as O He$\alpha$, O Ly$\alpha$, the Fe L complex, Ne He$\alpha$, Mg He$\alpha$, and Si He$\alpha$, while S line is not seen here. 
 
Firstly, we fitted each spectrum by single-component non-equilibrium ionization (VNEI) model. 
We employed \textbf{wabs} \cite{Morrison83} and \textbf{vnei} (NEI ver.2.0; Borkowski et al. 2001) in XSPEC version 12.3.1 \cite{Arnaud96}. 
In the model, the abundances of O, Ne, Mg, Si, and Fe were free while we set the abundances of C and N equal to O, S equal to Si, Ni equal to Fe, and other elements fixed to their solar values \cite{Anders89}.  
Other parameters were all free such as the electron temperature $kT_e$, the ionization timescale $\tau$ (a product of the electron density and the elapsed time after the shock heating), and the emission measure (EM $= \int n_e n_H dl$, where $n_e$ and $n_H$ are the number densities of hydrogen and electrons and $dl$ is the plasma depth). We also set the column density $\textit{N}_H$ free.  From the best-fit parameters, we found that the value of $kT_e$ ($\sim$0.4\,keV) is higher than that of the result at the NE rim obtained from \textit{Suzaku} observations ($\sim$0.2\, keV) \cite{Katsuda08-1}. 
Also the metal abundances such as Si ($\sim$1.0) and Fe ($\sim$0.4) show about two times higher values than those of the NE rim. 
These facts suggest that the X-ray emission in Pos-8 and 9 mainly consists of the high-$kT_e$ component which was explained in section \ref{sec:intro}. However, the values of reduced $\chi^2$ are 6.9 and 3.6 in Pos-8 and 9, respectively. The model is not enough to fit the data due to the simplicity of the model. Therefore, we intended to add the extra component to the VNEI model.

From the standpoint of the SNRs evolution, the X-ray emission from the SNRs have the two different origin. 
The blast wave from the supernova explosion sweeps the ambient medium, while the reverse shock propagates into the ejecta. 
Each shock wave heats up the swept-up ISM and the ejecta respectively. 
The shock-ISM interaction also produces the reflected shock which moves back through  previously swept up ISM \cite{Hester94}. 
Because of this, the X-ray spectra of the evolved SNR such as the Cygnus Loop should have a complicated structure. 
From the earlier observation of the NE to the SW regions, TKNM07 proposed the plasma structure of the Cygnus Loop as follows: the high-$kT_e$ ejecta component is surrounded by the low-$kT_e$ ISM component. 
They found that the spectra from most regions of the Cygnus Loop consist of the two-component VNEI model.
Thus we also employed two-component VNEI model which has two different electron temperatures. We found that this model cannot reach the physically meaningful results by setting all the parameters free. Therefore, in the low-$kT_e$ component, we fixed the metal abundances to the values from the NE rim observations \cite{Uchida06}. Other parameters were set free such as $kT_e$, $\tau$, and EM. In the high-$kT_e$ component, we set all parameters to those of the single-$kT_e$ VNEI model as explained in the above paragraph. As a result, the values of the reduced $\chi^2$ remain almost unchanged: 6.9 to 6.5 and 3.6 to 3.4 in Pos-8 and 9, respectively. 
 These large values are due to the fact that we took the spectra summed over the entire FOV in which there is a lot of structure. Therefore we divided our FOV into several regions for the spectral analysis. Although we employed the constant temperature, plane-parallel shock plasma model, \textbf{vpshock} instead of \textbf{vnei}, the best-fit parameters were almost unchanged and the values of the reduced $\chi^2$ were not significantly improved.

\subsection{Spatially Resolved Spectral Analysis}\label{sec:analysis}
\begin{figure}
  \begin{center}
    \includegraphics[width=55mm,angle=0]{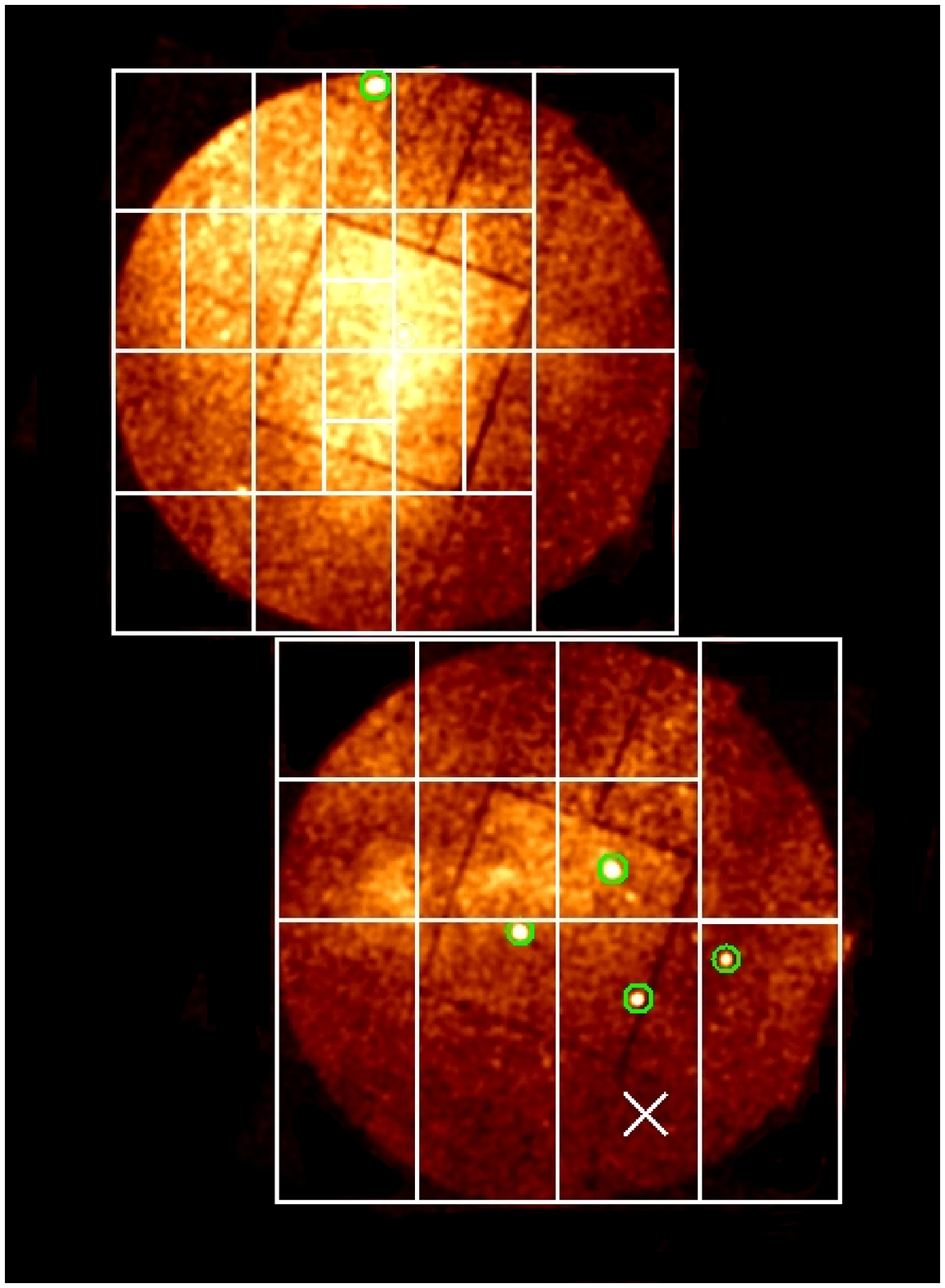} 
    \includegraphics[width=55mm,angle=0]{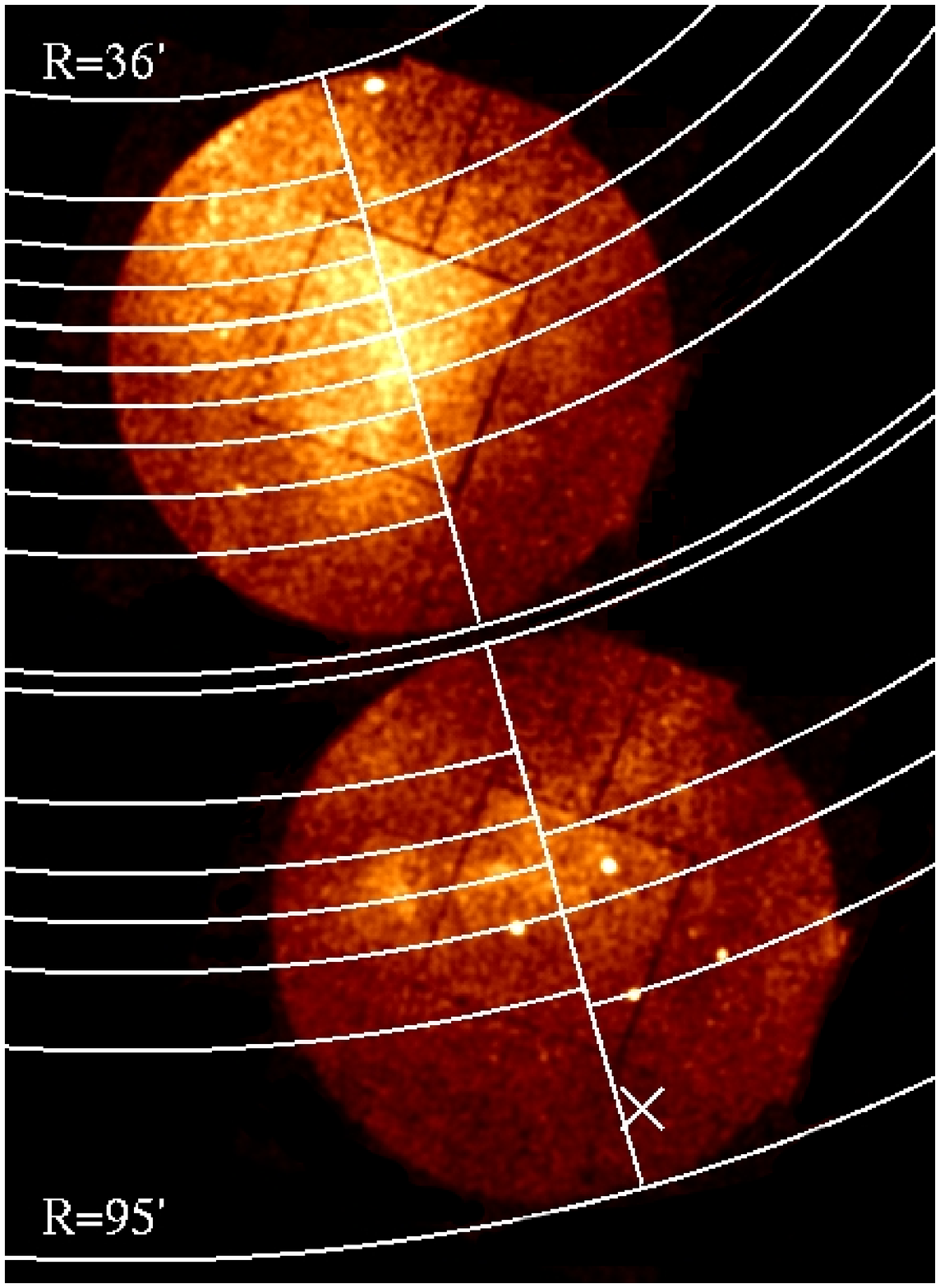} 
  \end{center}
  \caption{\textit{Left}: \textit{XMM-Newton} MOS broad-band image for the 0.3-3\,keV range. White lines represents the spectral extraction regions. The green circles show the point-like source regions excluded from our spectral analysis. The white X shows the center of the G72.9-9.0 estimated by Uyaniker et al. (2002) \textit{Right}: Same as the left panel, but for the different spectral extraction regions.}\label{fig:MOS}
\end{figure}

From Fig. \ref{fig:MOS}, we can see a lot of structures within Pos-8 and 9. For example, there is a region of high surface brightness at each center of Pos-8 and 9 even after correcting the vignetting effect. We notice that they are different in color in Fig. \ref{fig:HRI} right, which shows that the plasma temperatures are different from each region.  In order to investigate the detail plasma structure, we divided our FOV into a number of box regions for the spectral analysis. To equalize the statistics, we determined the box sizes such that each region has 7,500-15,000 photons for MOS 1 and 2. In this way, we have 33 regions (22 and 11 regions in Pos-8 and Pos-9, respectively). Figure \ref{fig:MOS} left panel shows the \textit{XMM-Newton} MOS broad band image for the 0.3-3.0 keV range and box regions are shown in white lines. 

To examine the plasma structure of Pos-8 and 9, we fitted 33 spectra extracted from box regions by the single-$kT_e$ VNEI model and two-$kT_e$ VNEI model, respectively. In the two-$kT_e$ VNEI model, we fixed the metal abundances of the low-$kT_e$ component to the result from the observations of the NE rim as explained in section~\ref{sec:vnei}.  As in the case of the fit for each whole region, the values of the reduced $\chi^2$ are improved $\sim$1.6 to $\sim$1.3 and $\sim$1.4 to $\sim$1.1 in Pos-8 and 9, respectively. The F-test probability ($ > 99\%$) shows that it is reasonable to add the extra low-$kT_e$ VNEI model in more than half of the regions. The best-fit parameters are shown in Figure \ref{fig:map} as the maps of the best-fit parameters. The averaged temperature of high- and low-$kT_e$ component are $\sim$0.4\,keV and $\sim$0.2\,keV, respectively. However, the low-$kT_e$ temperatures are determined only as the upper limit in several regions where the contribution of the low-$kT_e$ component is quite low as shown in the EM$\rm{_L}$ map in Fig. \ref{fig:map}. 

We compared these parameters and EMs of heavy elements with the results of Katsuda et al. (2008b). They observed the Cygnus Loop in seven pointings from the NE to the SW with \textit{Suzaku} and showed the best-fit parameters using the two-$kT_e$ VNEI model. One of their observation regions (named P16) is next to the NE part of Pos-8 (see Fig. \ref{fig:HRI}). From the results of P16 observations, the temperature of the high- and low-$kT_e$ component are 0.4\,keV and 0.2\,keV. These values are similar to our results.  Katsuda et al. concluded that the emission of the high-$kT_e$ component comes from the ejecta of the Cygnus Loop. Then we compared the EMs of O [=C=N], Ne, Mg, Si [=S], and Fe [=Ni] between in P16 and our FOV as shown in Table \ref{tab:EM}. Katsuda et al. (2008b) showed that each EM in their FOV reduces from the center to the outer region of the Loop. From Table \ref{tab:EM}, we found that this trend is also seen from P16 to our FOV.

Then, we determined the spectral extraction regions in different way to investigate the plasma structure from the inner side of the Loop to the outside. 

We divided our FOV into two paths: the east path and west path. Then we divided several annular sectors as shown in Fig. \ref{fig:MOS} right. To compare our analysis with that of TKNM07, we set the annular center on $20^{h}51^{m}34.7^{s},31^{\circ}00^{'}00^{''}$ (J2000).
In order to equalize the statistics, we determined the annular widths such that each sector has at least $\sim$10,000 $\pm 1,000$ photons.
In this way, we have 26 annular sectors (16 and 10 sectors in the east and the west path, respectively) whose angular distances from the center are from 35 arcmin to 95 arcmin. The width ranges from 1 arcmin to 6.5 arcmin. 

\begin{figure}
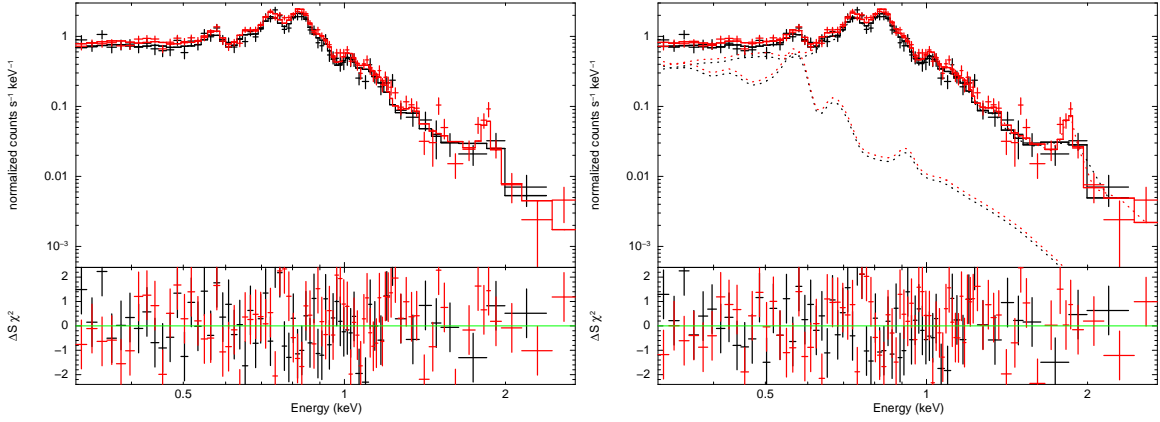

  \begin{center}
    \includegraphics[width=55mm,angle=-90]{f5a.eps}
     \includegraphics[width=55mm,angle=-90]{f5b.eps}
  \end{center}
  \caption{Example spectra at $R=$42.5$'$. The solid line of each panel shows the best-fit curve with the single-$kT_e$ VNEI model and the two-$kT_e$ VNEI model, respectively. Each lower panel shows the residual. The dotted lines of the right panel show individual component of the two-$kT_e$ VNEI model. }\label{fig:2comp}
\end{figure}

Figure \ref{fig:2comp} shows the example of the spectrum from the sector at $R=$ 42.5$'$, where $R$ represents the angular distance from the center. 
The left and right panels show the best-fit curves with the single- and the two-$kT_e$ VNEI model, respectively. The fitting parameters are set as explained in section~\ref{sec:vnei}.  Dotted lines in Fig. \ref{fig:2comp} represent the individual model. From Fig. \ref{fig:2comp} right, we found that the contribution of the low-$kT_e$ component is lower than that of the high-$kT_{e}$ component. The best-fit parameters are shown in Table \ref{tab:1comp} and \ref{tab:2comp}, respectively. The F-test probability ($ > 99\%$) shows that it is reasonable to add the extra low-$kT_e$ VNEI model in this sector.

\begin{figure}
  \begin{center}
    \includegraphics[width=90mm,angle=-90]{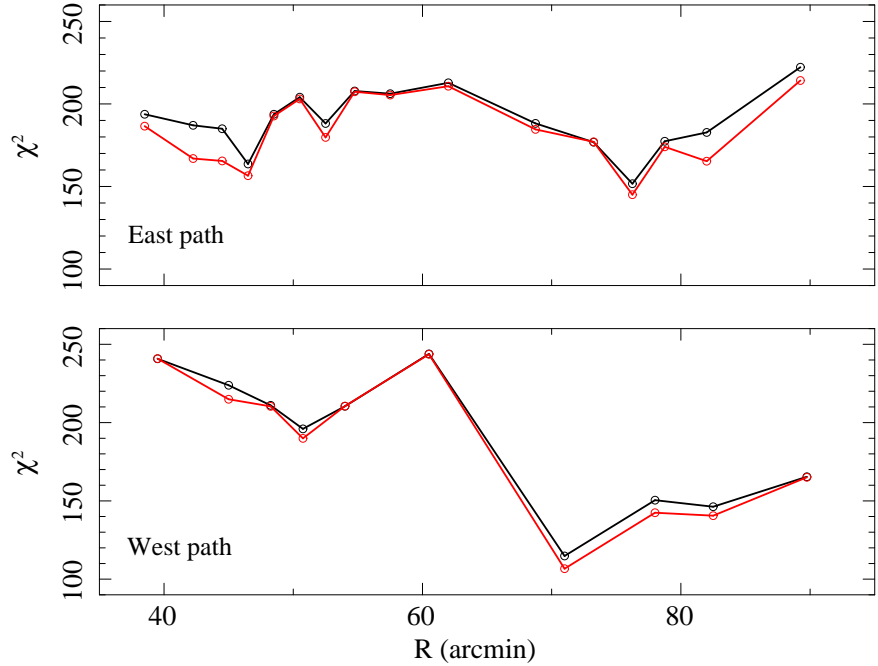}
  \end{center}
  \caption{Radial plot of the values of $\chi^2$ as a function of $R$ along the east path (top) and the west path (bottom). The single- and two-$kT_e$ VNEI model are shown in black and red, respectively. The degrees of freedom are all $\sim$130.}\label{fig:chi}
\end{figure}

Then, we analysed all other sectors in the same way. Figure \ref{fig:chi} shows the radial plot of the values of $\chi^2$ along the east path (top) and the west path (bottom). 
The single-$kT_{e}$ VNEI model is shown in black, while the two-$kT_{e}$ VNEI model is shown in red. From the results, we calculated the F-test probability and determined whether or not the extra component is needed for each sector. Applying the significance level of 99$\%$, the extra component is not required at $47.5' < R < 75.0'$, and $77.5' < R < 95.0'$ along the east path, and  $36.0' < R < 43.0'$, $47.0' < R < 65.0'$, and $85.0' < R < 95.0'$ along the west path. In other words, $\sim$60$\%$ of our FOV requires the two-$kT_e$ VNEI model. We find that the fit shown in Fig.\ref{fig:2comp} (at $R=$ 42.5$'$) is improved the most by using the two-$kT_e$ VNEI model. 
Even in this sector, the contribution of the additional low-$kT_{e}$ component is not so large. 

\section{Discussion}
\begin{figure}
  \begin{center}
    \includegraphics[width=45mm,angle=0]{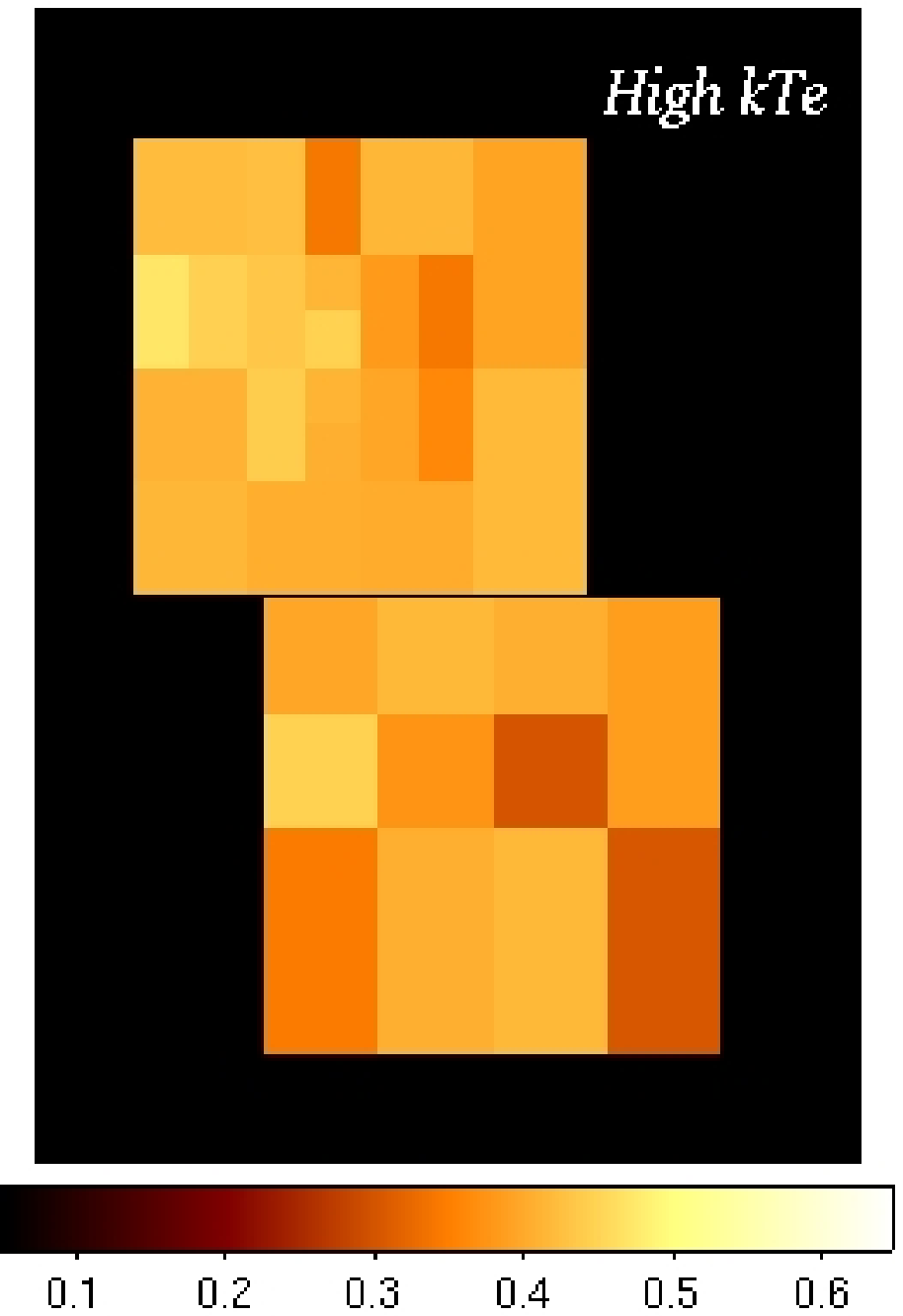}
    \includegraphics[width=45mm,angle=0]{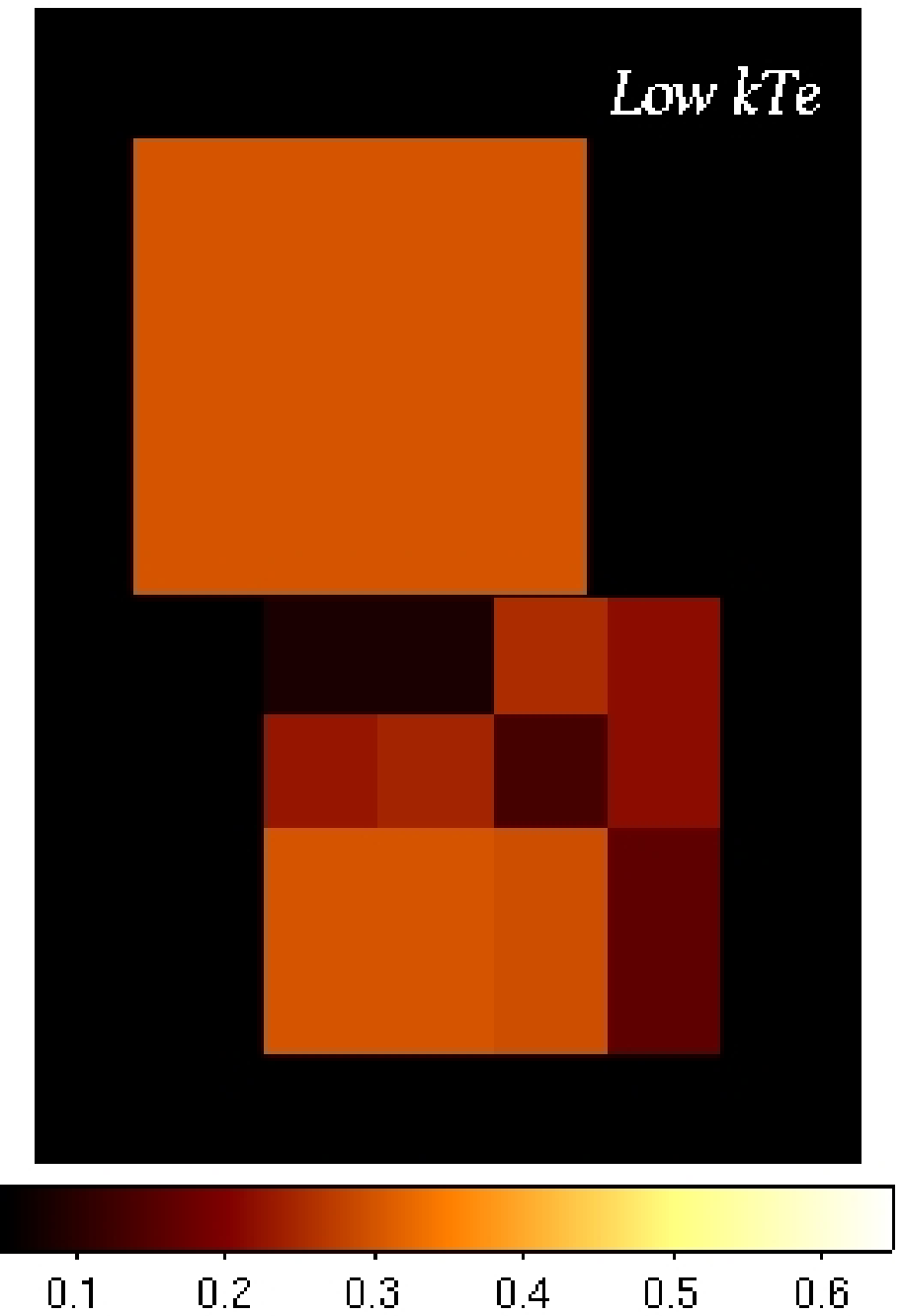}
    \includegraphics[width=45mm,angle=0]{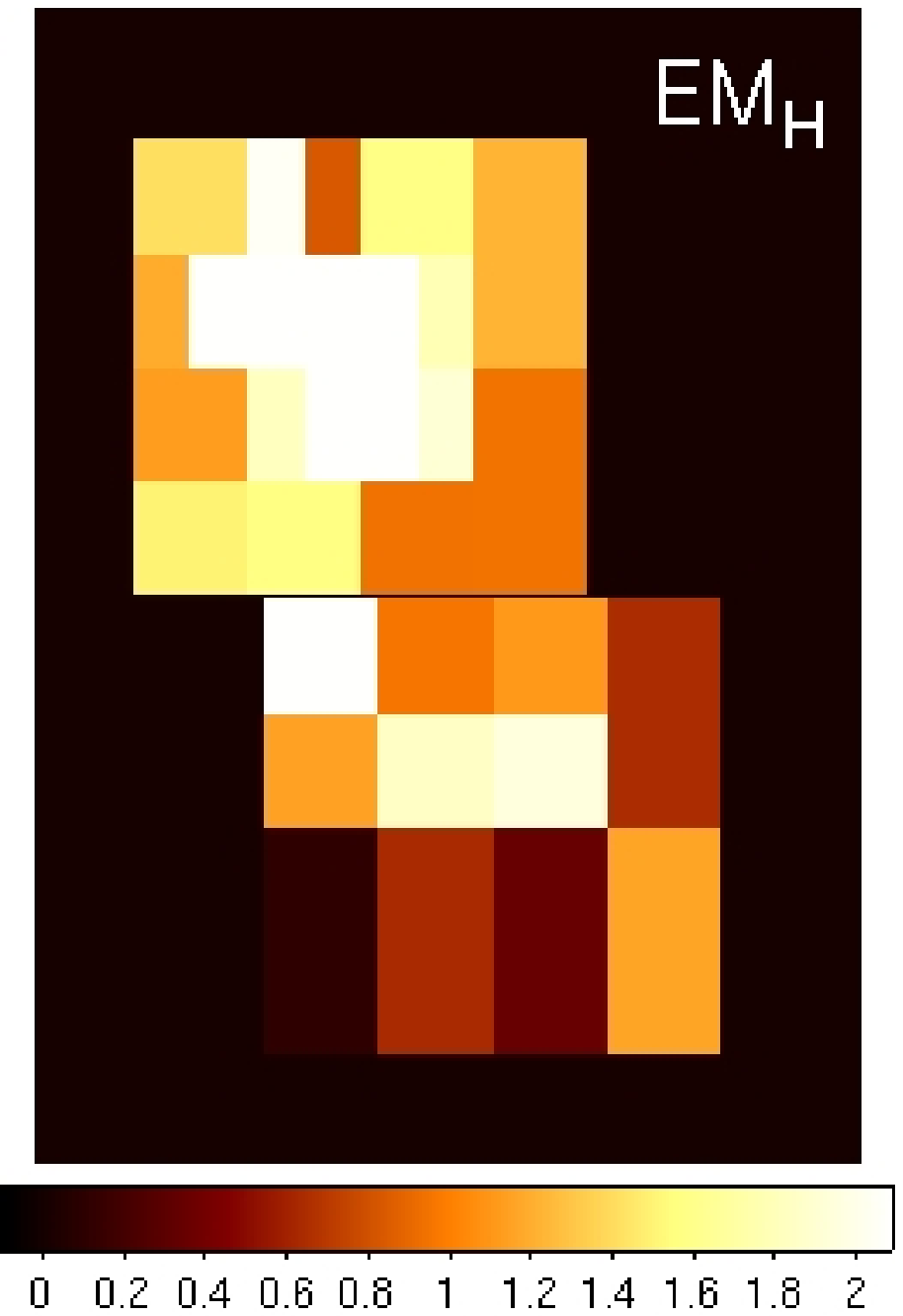}
    \includegraphics[width=45mm,angle=0]{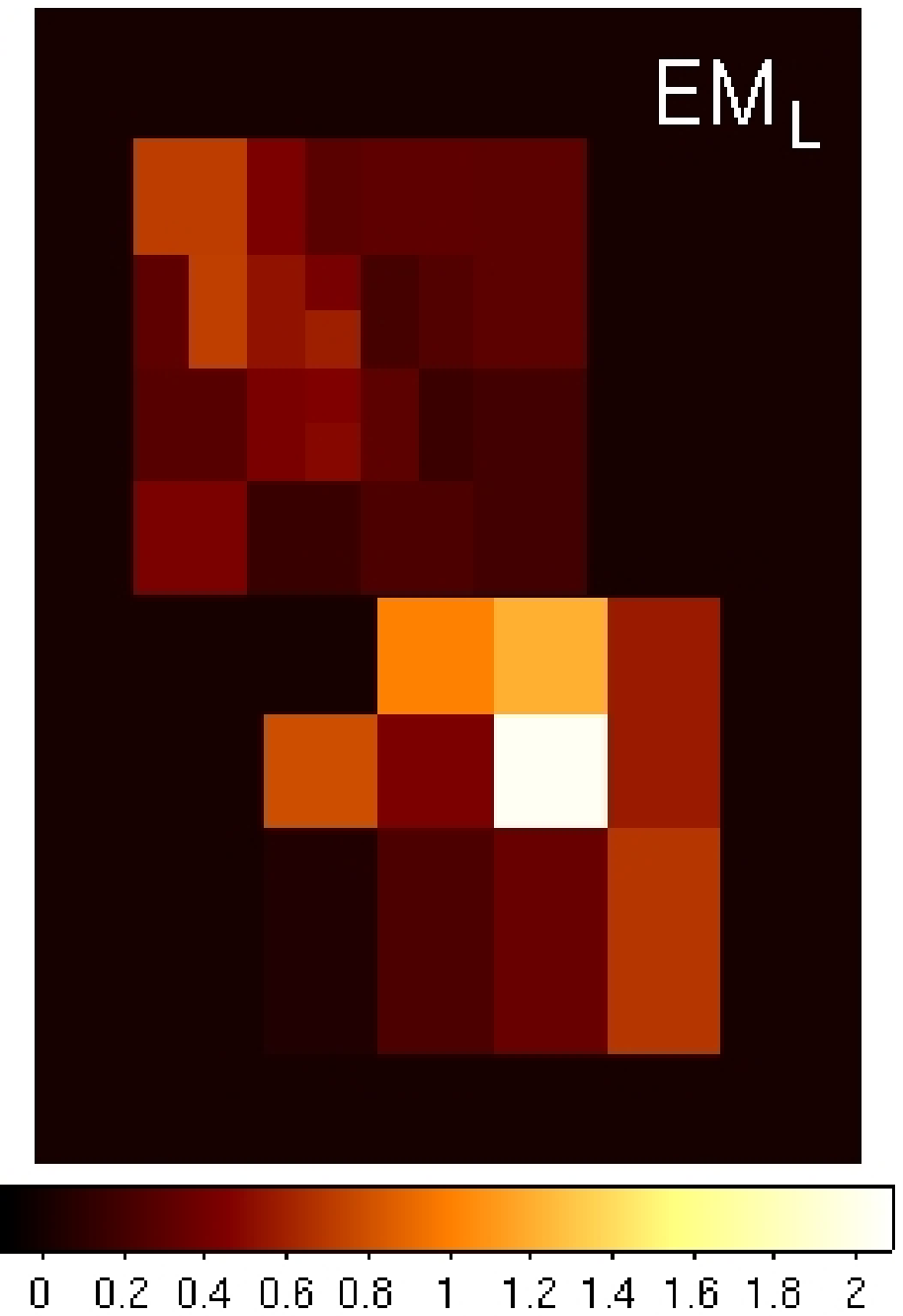}
    \includegraphics[width=45mm,angle=0]{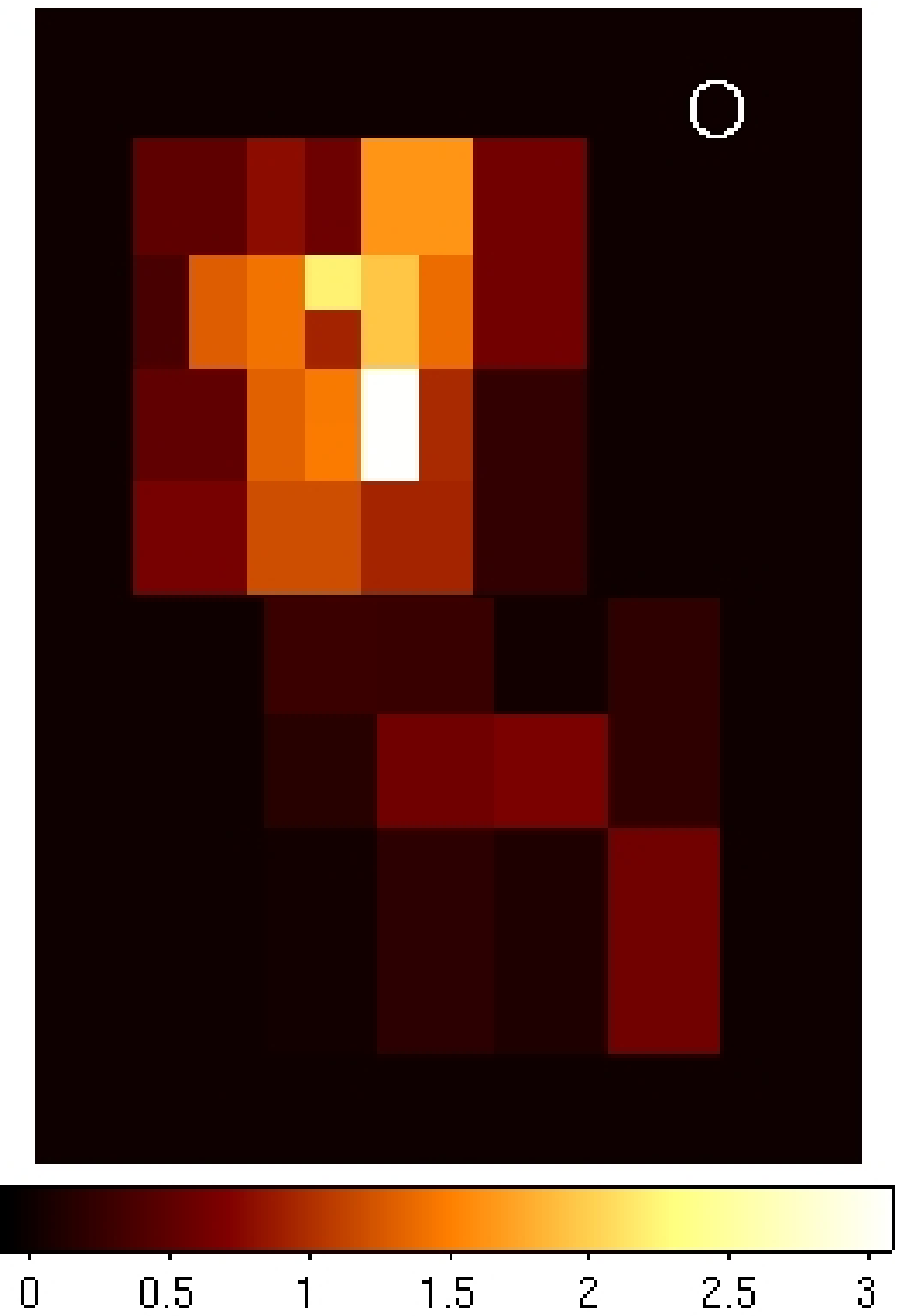}
    \includegraphics[width=45mm,angle=0]{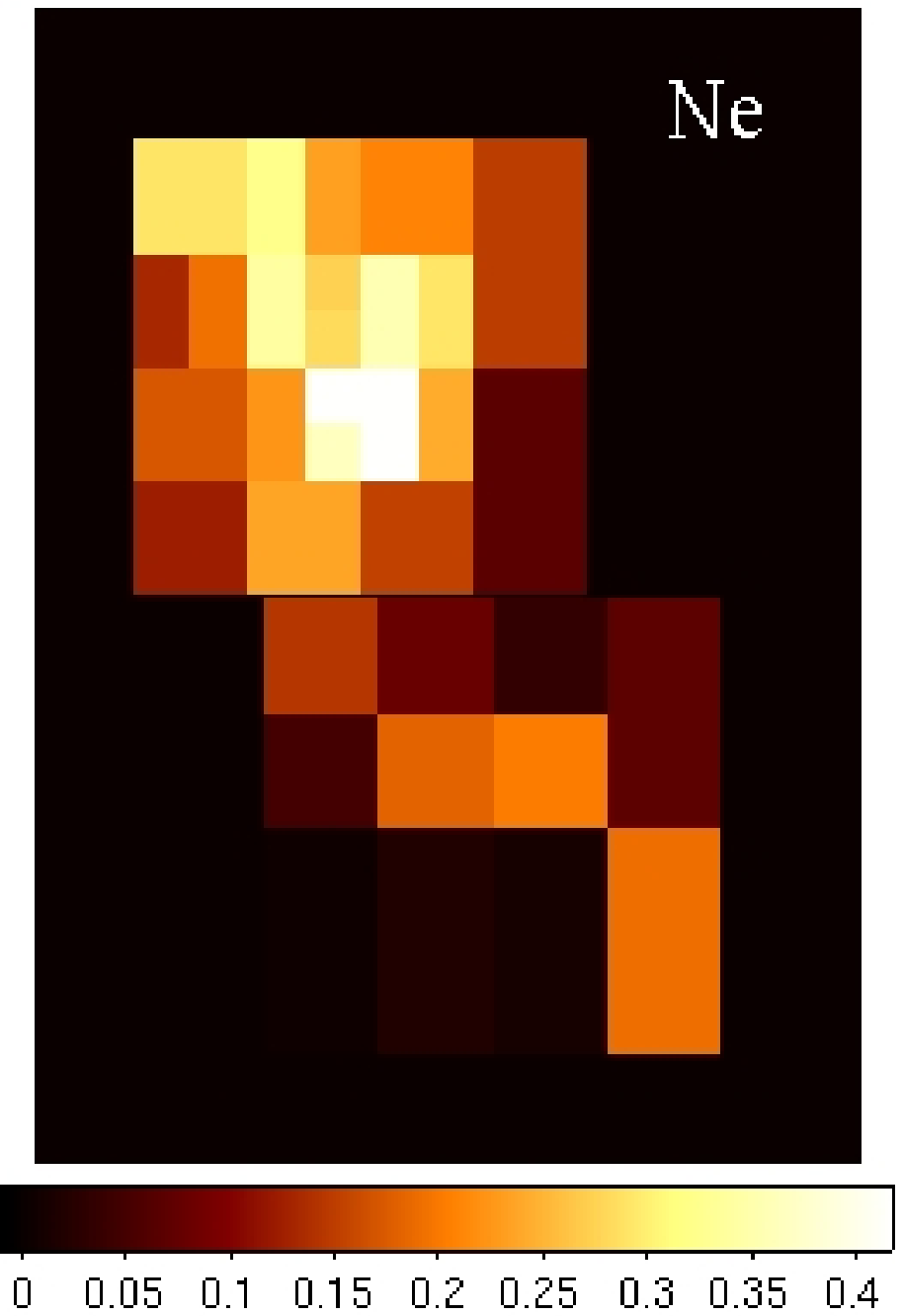}
    \includegraphics[width=45mm,angle=0]{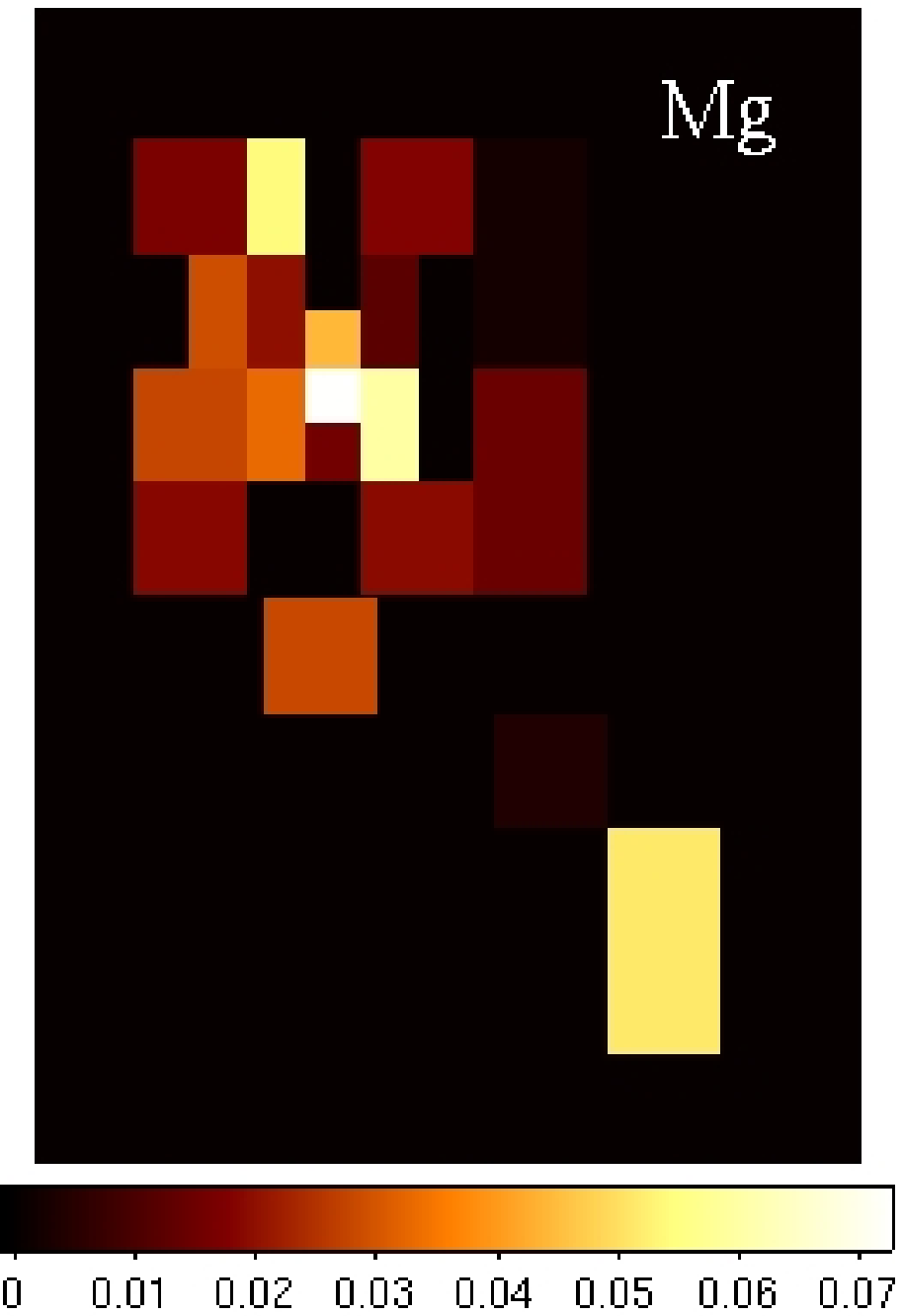}
    \includegraphics[width=45mm,angle=0]{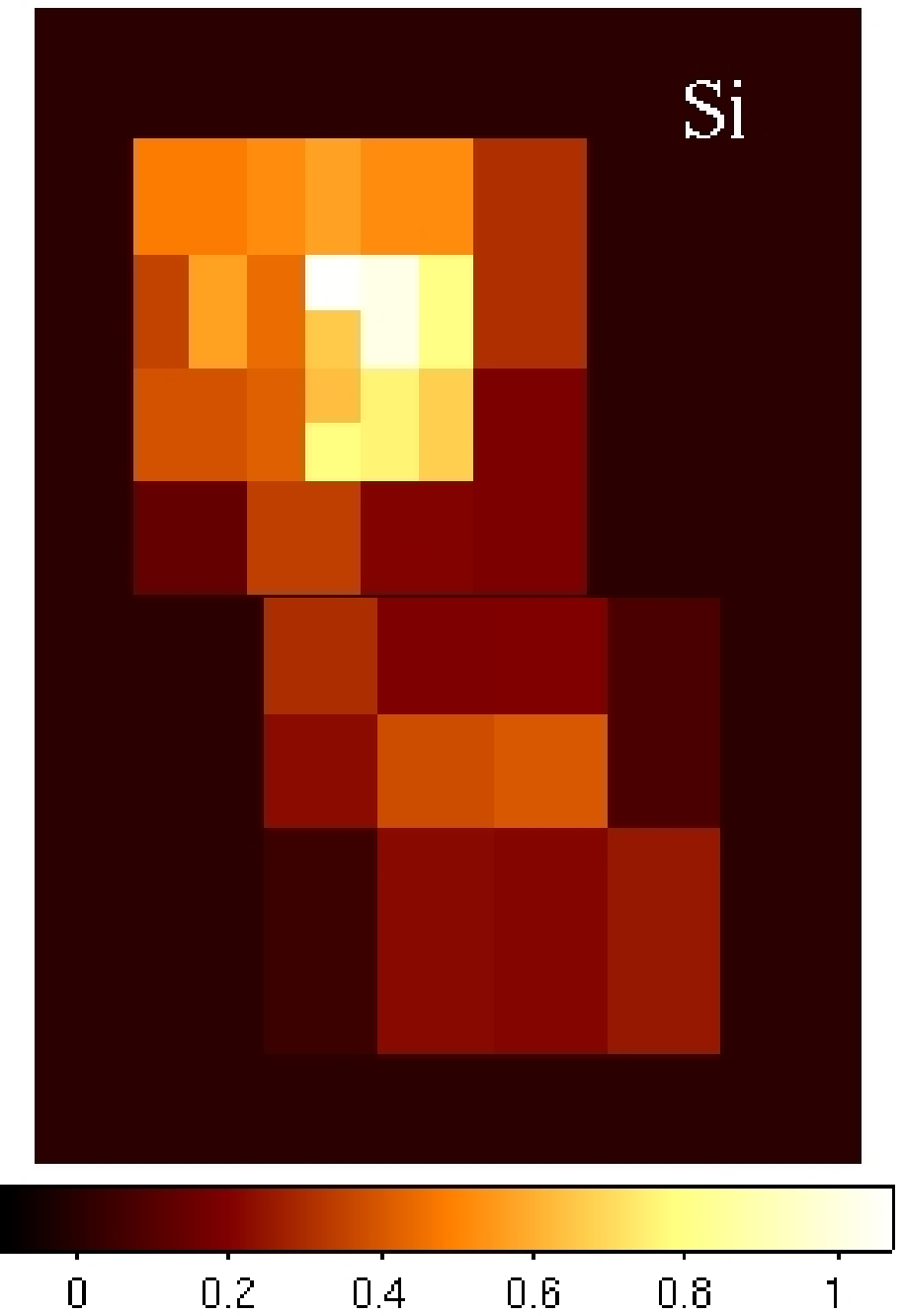}
    \includegraphics[width=45mm,angle=0]{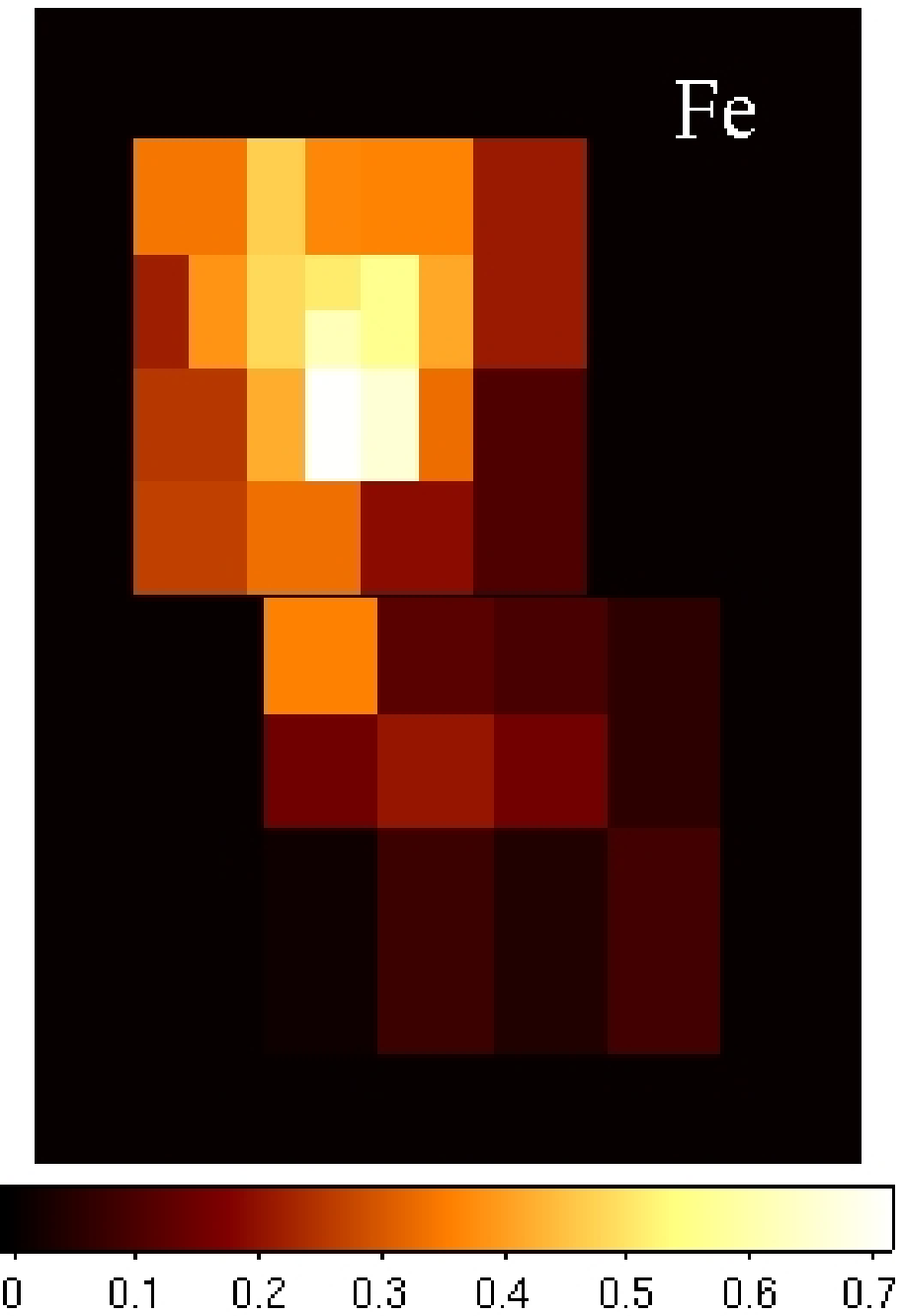}
 \end{center}
  \caption{Maps of the best-fit parameters. EM$\rm{_H}$ and EM$\rm{_L}$ mean the emission measure of the high- and low-$kT_e$ component, respectively. Last five panels show the EMs of O [=C=N], Ne, Mg, Si [=S], and Fe [=Ni] for the high-$kT_e$ component in units of 10$^{14}$cm$^{-5}$. The values of $kT_e$ and EM$_{[H,L]}$ are in units of keV and 10$^{18}$cm$^{-5}$, respectively. }\label{fig:map}
\end{figure}

\begin{figure}
  \begin{center}
    \includegraphics[width=110mm,angle=0]{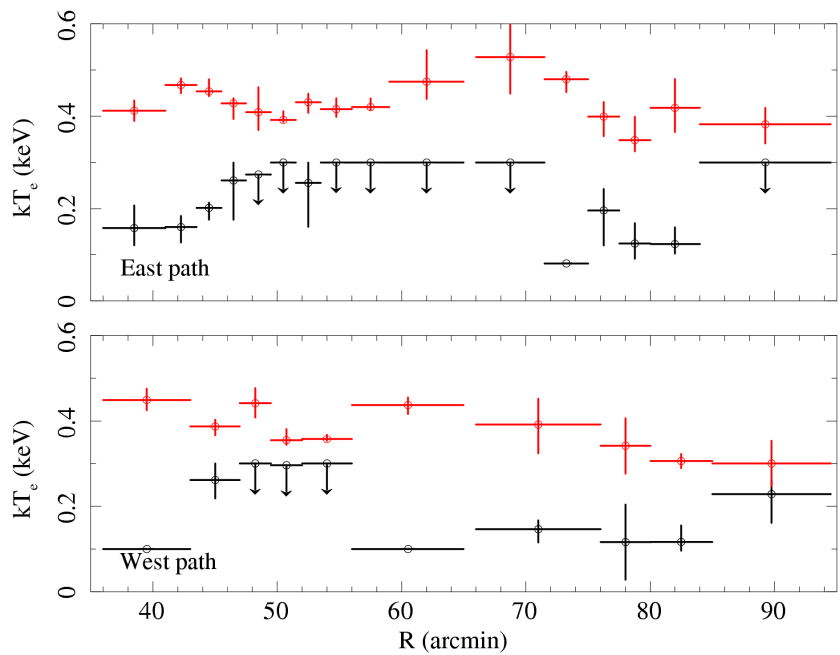}
  \end{center}
  \caption{Temperature distributions of the two components as a function of $R$ along the east path (top) and the west path (bottom). Red shows the high-$kT_e$ component, while black shows the low-$kT_e$ component.}\label{fig:kTe}
\end{figure}

The first two panels of Fig. \ref{fig:map} show the temperature distributions of the two components based on the analysis in Fig. \ref{fig:MOS} left. Figure \ref{fig:kTe} shows the temperature distributions along the east path (top) and the west path (bottom) based on the analysis in Fig. \ref{fig:MOS} right. Black and red represent the low-$kT_{e}$ temperature and the high-$kT_{e}$ temperature. From Fig. \ref{fig:map} and \ref{fig:kTe}, the averaged values of low- and high-$kT_{e}$ temperature are $\sim$0.2\,keV and $\sim$0.4\,keV, respectively. In this way, we clearly separated the high-$kT_{e}$ component and the low-$kT_{e}$ component just as the observation obtained in Katsuda et al. (2008b) and TKNM07. 
TKNM07 showed the temperature of the low-$kT_{e}$ component is almost constant ($\sim$0.2\,keV) along the diameter, while that of the high-$kT_{e}$ component is different in NE ($\sim$0.6\,keV) and SW ($\sim$0.4\,keV). Since our FOV is very close to SW, our result shows smooth extrapolation from that of TKNM07 in SW rather than that in NE. 

\begin{figure}
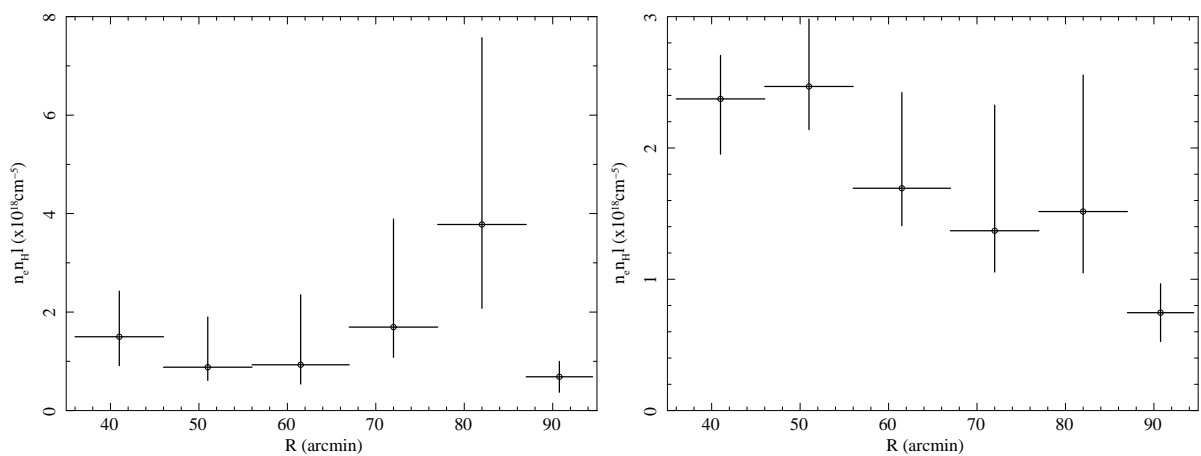

  \begin{center}
    \includegraphics[width=60mm,angle=-90]{f8a.eps}
    \includegraphics[width=60mm,angle=-90]{f8b.eps}
  \end{center}
  \caption{\textit{Left}: EM$\rm{_L}$ distribution as a function of $R$. \textit{Right}: Same as the left, but for EM$\rm{_H}$.}\label{fig:EMnorm}
\end{figure}

\begin{figure}
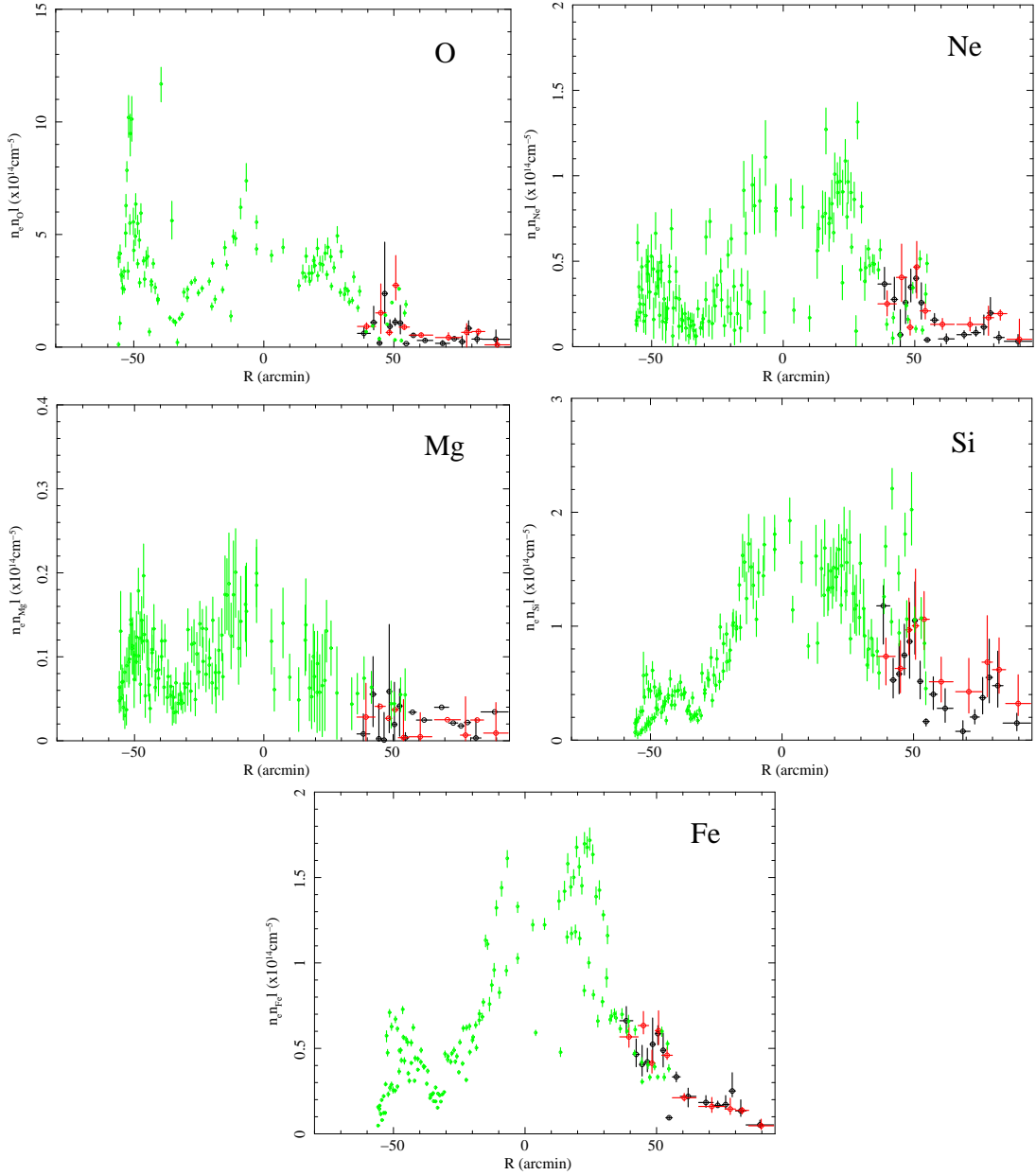

  \begin{center}
    \includegraphics[width=55mm,angle=-90]{f9a.eps}
    \includegraphics[width=55mm,angle=-90]{f9b.eps}
    \includegraphics[width=55mm,angle=-90]{f9c.eps}
    \includegraphics[width=55mm,angle=-90]{f9d.eps}
    \includegraphics[width=55mm,angle=-90]{f9e.eps}
   \end{center}
  \caption{EM distributions for various metals (O [=C=N], Ne, Mg, Si [=S], and Fe [=Ni]) in the ejecta. Black and red show the west path and east path, respectively. Green shows the result of TKNM07 taken from Pos-2 to Pos-6.}\label{fig:EM}
\end{figure}

The third and fourth panels of Fig. \ref{fig:map} shows the EM distributions for each component. Although there are some structures seen in the EM of the high-$kT_{e}$ component (EM$\rm{_H}$) map, it is clear that the EM$\rm{_H}$ is higher in Pos-8 than that in Pos-9. The EM of the low $kT_{e}$ component (EM$\rm{_L}$) in all of our FOV are lower than those of EM$\rm{_H}$. Figure \ref{fig:EMnorm} shows the radial profile of the EMs for each component.
In this figure, we calculated the EMs as a function of $R$ into 10\,arcmin bin. The EM$\rm{_L}$ stays almost constant while it peaks around $R=80'$. From the morphological point of view, the Cygnus Loop has an almost circular shape with a radius $\sim80'$ except for the south ``blow-out''. Then, it is suggested that the EM$\rm{_L}$ distribution reflects the rim brightening structure around the ``blow-out'' region. On the other hand, the value of EM$\rm{_H}$ gradually decreases from the center to the outer region. 
This decrease can be easily explained by assuming that the emission comes from the ejecta component filling inside the Cygnus Loop. Then we also measured the EMs of various heavy elements in the high-$kT_{e}$ component such as O [=C=N], Ne, Mg, Si [=S], and Fe [=Ni] and compared them with the result of TKNM07. Figure \ref{fig:EM} shows the EM distribution for these elements as a function of $R$. We also plot the results of Pos-2 to 6 (TKNM07) in the same panels. 
Although some structures are remaining in the annular regions as seen in Fig. \ref{fig:MOS}, the radial distribution of each EM clearly shows the smooth extrapolation of TKNM07's result. They showed the decrease of each EM from the center to the outer region and concluded that the high-$kT_{e}$ component is derived from the Cygnus Loop ejecta. It is reasonable to understand that the EMs in Pos-8 and 9 shows a smooth connection to those in their FOV. Therefore we concluded that the high-$kT_{e}$ component originates from the ejecta of the Cygnus Loop.

\begin{figure}
  \begin{center}
    \includegraphics[width=80mm,angle=-90]{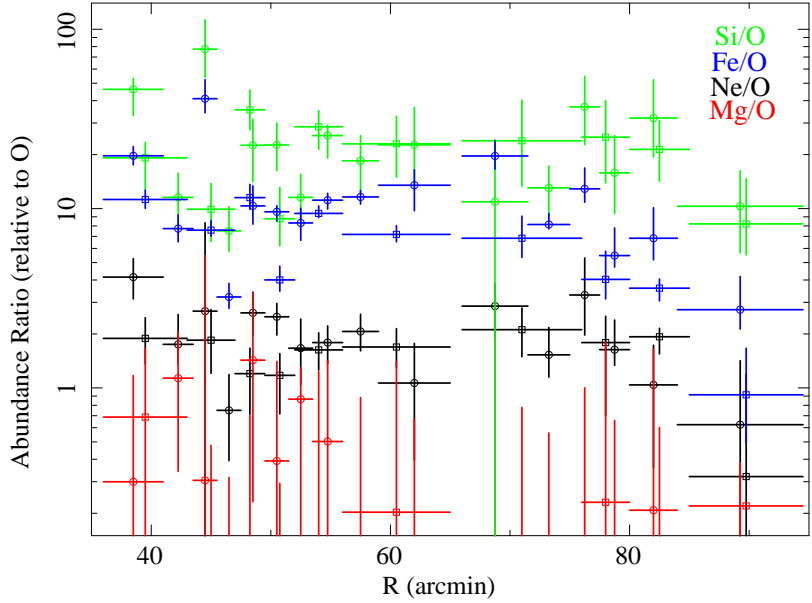}
  \end{center}
  \caption{Distributions of relative abundance of heavy elements to O are shown as a function of $R$. Si/O, Fe/O, Ne/O, and Mg/O are shown in green, blue, black, and red, respectively. The results of east path and those of west path are plotted in the same color.}\label{fig:abundance_ratio}
\end{figure}

From the fitting parameters of the high-$kT_{e}$ component, we calculated the abundances of ejecta component for various elements. Figure \ref{fig:abundance_ratio} shows the abundance ratios of heavy elements (Ne: black, Mg: red, Si [=S]: green, Fe [=Ni]: red) relative to O. From Fig. \ref{fig:abundance_ratio}, we found that Si/O ($\sim$20) and Fe/O ($\sim$10) are heavily over abundant and Ne/O is $\sim$2, while Mg/O ($<$1) is depleted. This tendency is kept throughout our observing region. The other observations of the ejecta in the Loop such as TKNM07 and Katsuda et al. (2008b) showed the similar results. 

Uyaniker et al. (2002) and Sun et al. (2006) reported that the Cygnus Loop consists of two SNRs interacting with each other in the SW. Their main arguments are the difference of the radio morphology and the polarization intensity between the main part of the Cygnus Loop and the south ``blow-out'' region. However, based on the X-ray data, we found that there is no evidence of the extra SNR within our FOV. If these SNRs are at the same distance as claimed by Uyaniker et al. (2002) and Sun et al. (2006), the smaller radius of the extra SNR, $\sim$7\,pc ($R$/$\sim$0.7$^\circ$)($d$/540\,pc), than that of the Cygnus Loop, $\sim$13\,pc ($R$/$\sim$1.4$^\circ$)($d$/540\,pc), strongly suggests that the extra SNR is younger than the Cygnus Loop is. If we employ the Sedov-Taylor solution, the temperature $T$ of the extra SNR is 

\begin{equation}
T \sim {\rm1.8\,keV} \left(\frac{{\rm E_{0}}}{\rm 10^{51}\,ergs}\right)\left(\frac{n}{\rm1\,cm^{-3}}\right)^{-1}\left(\frac{R}{\rm7\,pc}\right)^{-3}
\end{equation}

where E$_{0}$ and $n$ are the explosion energy and the surrounding medium density of the extra SNR, respectively. Therefore, the temperature of the extra SNR should be significantly higher than that of the Cygnus Loop.
However, we found no sign of such high temperature plasma. 
The spectra from all regions are almost represented by a single-$kT_{e}$ VNEI model ($\sim$0.4\,keV). 
If we add an extra component, we found in section~\ref{sec:specana} that the extra component shows low temperature rather than high temperature. 
Furthermore, from Fig. \ref{fig:kTe}, the temperatures of each component are in good agreement with those obtained in other regions of the Loop (TKNM07; Katsuda et al. 2008b). 
This result suggests that the X-ray emission from the SW region mainly comes from the  Cygnus Loop. 
If the secondary SNR exists in the SW at the same distance, the contribution of the X-ray emission to the spectra is much less than that of the Cygnus Loop. 
We cannot rule out the possibility that the extra SNR exists far side of the Cygnus Loop. 
However, even if that is the case, the fact remains that the spectrum from our FOV mainly consists of the ejecta and the ISM components of the Cygnus Loop. 
As a result, we find no evidence in X-ray that there exists the second SNR at the same distance to the Cygnus Loop. 
 
\begin{figure}
  \begin{center}
    \includegraphics[width=130mm,angle=0]{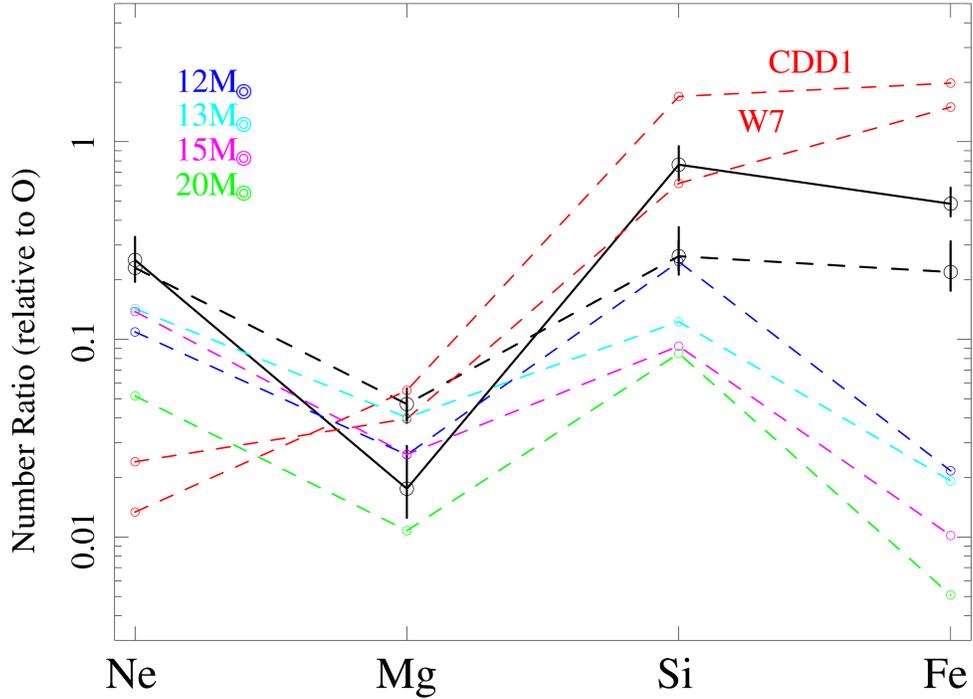}
  \end{center}
  \caption{Number ratios of Ne, Mg, Si, and Fe relative to O of the high-$kT_e$ component, estimated for the entire Loop (\textit{solid line}). Dotted and solid black lines show the result from TKNM07's FOV and that from our FOV. Dotted red lines represent the CDD1 and W7 Type Ia supernova models of Iwamoto et al. (1999). Dotted blue, light blue, magenta, and green lines represent core-collapse models with progenitor masses of 12, 13, 15, 20 M$_\odot$, respectively (Woosley \& Weaver 1995).}\label{fig:Number}
\end{figure}

   Then, we can estimate the mass of the progenitor star of the Cygnus Loop from the EMs of the high-$kT_{e}$ component, assuming that all these emissions come from the Loop. Then, we multiplied the EMs by the area of each annular sector and integrated the EMs along the path. In this way, we obtained the emission integral (EI $= \int n_e n_X dV$, where $dV$ is the X-ray-emitting volume) for O, Ne, Mg, Si, and Fe. Table \ref{tab:EI} shows the calculated EI of each element. To compare our data with the supernova explosion models, we calculated the ratios of Ne, Mg, Si and Fe relative to O. Figure \ref{fig:Number} shows the number ratios of Ne, Mg, Si and Fe relative to O of the ejecta component. We also plotted the result from TKNM07, the core-collapse models \cite{Woosley95} for various progenitor masses and Type Ia supernova models \cite{Iwamoto99} for comparison. The type Ia supernova yields more Fe than our results but less Ne. On the contrary, the number ratios of Si and Fe are higher than those of any core-collapse models, which attributes to the fact that our calculations are derived from the small part of the Cygnus Loop. These models were calculated under the assumption of the symmetric explosion. However, TKNM07 and Katsuda et al. (2008b) reported the asymmetry of the metal distribution of the ejecta component: Si and Fe were more abundant, while Ne and Mg were less abundant in the SW rather than that in the NE. These results support an asymmetric explosion of the progenitor star. Our FOV is close to the SW region in TKNM07 rather than the NE region. Taking into account the effect of the asymmetric structure, we support the idea that the Cygnus Loop originates from the core-collapse explosion rather than the Type Ia supernova. 

\section{Conclusion}
We observed the SW region of the Cygnus Loop with \textit{XMM-Newton}. To examine the plasma structure, we divided our FOV in two different ways: 33 box sectors and 26 annular sectors. We fitted the spectrum extracted from each region with two-$kT_{e}$ VNEI model. The plasma structure of the low-$kT_{e}$ component and that of the high-$kT_{e}$ component are quite different from each other: each temperature is $\sim$0.2\,keV and $\sim$0.4\,keV for the former and the latter, respectively.

The EM distribution of the low-$kT_{e}$ component suggest the rim brightening structure, while that of the high-$kT_{e}$ component monotonously decreases from the center of the Loop to the outside.  In the high-$kT_{e}$ component, the abundances of Si and Fe are relatively high compared to those of Ne and Mg. The distributions of EMs as well as the relative abundances in the high-$kT_{e}$ component match the view that the low- and high-$kT_{e}$ components, respectively, originate from the ISM and the ejecta of the Cygnus Loop, which was derived by earlier observations such as TKNM07 or Katsuda et al. (2008b).
We found that the emission from this ISM component is relatively weak. This suggests that the thickness of the shell is thin in Pos-8 and 9. 
We also calculated the relative abundances of Ne, Mg, Si, and Fe to O in the ejecta component for the entire FOV, and estimated the origin of the Cygnus Loop as the core-collapse explosion rather than the Type Ia supernova. 
We found no evidence in X-ray that the nature of the ``blow-out'' region originated from the extra SNR. 

\acknowledgments
We thank H. Kosugi's careful reading of the manuscript. This work is partly supported by a Grant-in-Aid for Scientific Research
by the Ministry of Education, Culture, Sports, Science and Technology
(16002004).  This study is also carried out as part of the 21st Century
COE Program, \lq{\it Towards a new basic science: depth and
synthesis}\rq. H.U. and S.K. are supported by JSPS Research Fellowship for Young
Scientists. 

\begin{table}
  \caption{Spectral Fit Parameters}\label{tab:1comp}
    \begin{tabular}{ll}
       \hline
      \hline
 Parameter & Pos-8 \\
      \hline
      N$\rm _H$ [10$^{20}$cm$^{-2}$] &  4 $\pm$ 1\\
      $kT_e$ [keV] & 0.46 $\pm$ 0.02 \\ 
      O(=C=N) & 0.04 $\pm$ 0.01 \\
      Ne & 0.09 $\pm$ 0.01 \\
      Mg & 0.04 $\pm$ 0.02 \\
      Si & 0.79 $\pm$ 0.16 \\
      Fe(=Ni) & 0.33 $\pm$ 0.03 \\
      log $\tau$ & 10.58 \\
      EM\tablenotemark{a} [10$^{18}$cm$^{-5}$] & 2.54 $\pm$ 0.56\\
      $\chi ^2$/dof & 187/133\\
      \hline
    \end{tabular}
\tablecomments{Other elements are fixed to solar values. The errors are in the range $\Delta\chi^2 < 2.7$ for one parameter.}
\tablenotetext{a}{ Emission Measure, $\int n_e n_H dl$}
\end{table}

\begin{table}
  \caption{Spectral Fit Parameters}\label{tab:2comp}
    \begin{tabular}{ll}
       \hline
      \hline
 Parameter & Value\\
      \hline
      N$\rm _H$ [10$^{20}$cm$^{-2}$] & $<$\,4 \\
      Low-$kT_e$ component: \\
      \ \ $kT_e$ [keV] & 0.26 $\pm$ 0.07  \\ 
      \ \ C & 0.27 (fixed)\\
      \ \ N &  0.10 (fixed)\\
      \ \ O &  0.11 (fixed)\\
      \ \ Ne &  0.21 (fixed)\\
      \ \ Mg &  0.17 (fixed)\\
      \ \ Si &  0.34 (fixed)\\
      \ \ S &  0.17 (fixed)\\
      \ \ Fe(=Ni) &  0.20 (fixed)\\
      \ \ log $\tau$ & 10.62  \\
      \ \ EM\tablenotemark{a} [10$^{18}$cm$^{-5}$] & $<$\,0.57\\
      High-$kT_e$ component: \\
      \ \ $kT_e$ [keV] & 0.47 $\pm$ 0.02  \\ 
      \ \ O(=C=N) & 0.06 $\pm$ 0.02 \\
      \ \ Ne & 0.08 $\pm$ 0.03 \\
      \ \ Mg & 0.06 $\pm$ 0.03 \\
      \ \ Si & 0.65 $\pm$ 0.14 \\
      \ \ Fe(=Ni) & 0.42 $\pm$ 0.05 \\
      \ \ log $\tau$ & 11.29 $\pm$ 0.03 \\
      \ \ EM\tablenotemark{a} [10$^{18}$cm$^{-5}$] & 1.88 $\pm$ 0.38\\
$\chi ^2$/dof & 167/130 \\
      \hline
    \end{tabular}
\tablecomments{Other elements are fixed to solar values. The errors are in the range $\Delta\chi^2 < 2.7$ for one parameter.}
\tablenotetext{a}{ Emission Measure, $\int n_e n_H dl$}
\end{table}

\begin{table}
\begin{center}
  \caption{Comparison between the averaged EM for each element in our FOV and P16 (Katsuda et al. 2008b)}\label{tab:EM}
    \begin{tabular}{lllll}
       \hline
      \hline
    EM of Each Element & & & \\
    ($10^{14}$ cm$^{-5}$) & P16 & Pos-8 & Pos-9 \\
      \hline
      O(=C=N) & 4.2 & 1.2 & 0.28 \\
      Ne      & 0.42 & 0.25 & 0.09 \\
      Mg      & 0.04 & 0.02 & 0.01  \\
      Si      & 1.7 & 0.54 & 0.22 \\
      Fe(=Ni) & 1.1 & 0.41 & 0.12  \\
      \hline
    \end{tabular}
\end{center}
\end{table}

\begin{table}
  \caption{Calculated Emission Integrals ($=\int n_e n_X dV$) of the Cygnus Loop Ejecta}\label{tab:EI}
  \begin{center}
    \begin{tabular}{ll}
       \hline
      \hline
      & \ \ \ \ \ \ EI \\
 Element & ($10^{52}$ cm$^{-3}$) \\
      \hline
      O & 1.34 $\pm$ 0.20\\ 
      Ne & 0.34 $\pm$ 0.03\\
      Mg & 0.02 $\pm$ 0.01\\
      Si & 1.02 $\pm$ 0.10\\
      Fe & 0.65 $\pm$ 0.03\\ 
      \hline
    \end{tabular}
  \end{center}
\end{table}

\clearpage

\end{document}